\documentclass[pdflatex,sn-mathphys-num]{sn-jnl}

\usepackage{graphicx}%
\usepackage{multirow}%
\usepackage{amsmath,amssymb,amsfonts}%
\usepackage{amsthm}%
\usepackage[title]{appendix}%
\usepackage{xcolor}%
\usepackage{textcomp}%
\usepackage{manyfoot}%
\usepackage{booktabs}%
\usepackage{url}
\usepackage{algorithm}%
\usepackage{algorithmicx}%
\usepackage{algpseudocode}%
\usepackage{listings}%
\usepackage{booktabs}

\theoremstyle{thmstyleone}%
\theoremstyle{thmstyletwo}%

\theoremstyle{thmstylethree}%

\begin{document}

\title[Article Title]{Magnetic field controlled nucleation and size selection of silver nanoparticles}


\author[1]{\fnm{Yazeed} \sur{Tawalbeh}}\email{100059491@ku.ac.ae}

\author*[1,2]{\fnm{Mauro} \sur{F. Pereira}}\email{mauro.pereira@ku.ac.ae}

\affil*[1]{\orgdiv{Physics Department}, \orgname{Khalifa University}, \orgaddress{\street{Shakhbout Bin Sultan St}, \city{Abu Dhabi}, \postcode{127788}, \country{United Arab Emirates}}}

\affil[2]{\orgdiv{Institute of Physics}, \orgname{Czech Academy of Sciences}, \orgaddress{\city{Prague}, \postcode{18221}, \country{Czech Republic}}}


\abstract{
We examine the reduction of silver nanoparticle (AgNP) size under an external magnetic field within a classical nucleation theory framework combined with a sphere-packing description of atomic assembly. The model incorporates magnetic free-energy contributions arising from the coupling between the applied field and the magnetic susceptibility of the nucleating material, yielding a closed-form relation between nanoparticle radius and field strength. Our approach reproduces the experimentally observed decrease in the most-probable particle radius from approximately $170\text{ nm}$ at $\mathcal{B}=49.27\text{ mT}$  when the magnetic field is oriented parallel to the stirring plane, and to $155\text{ nm}$ at $\mathcal{B}=180.78\text{ mT}$ in the perpendicular configuration. Across the investigated field range, the theoretical predictions remain consistent with experimental measurements obtained under continuous mechanical stirring, supporting the interpretation that the observed size reduction originates from a magnetic-field-induced modification of the nucleation free-energy landscape. Within the limits of classical capillarity and spherical demagnetization, the results provide a physically transparent and computationally efficient framework for understanding magnetic-field-controlled nanoparticle size selection.
}



\maketitle
\section{introduction} Nanoparticles (NPs) exhibit chemical, optical and physical properties that differ markedly from their bulk counterparts, largely due to quantum confinement effects and their high surface-to-volume ratio \cite{10.1016/j.jallcom.2023.172999, 10.1016/j.matpr.2023.05.456, doi.org/10.1016/j.ceramint.2024.06.067, 10.1007/s11051-024-06119-8,10.1016/j.totert.2023.100038, 10.1039/D5YA00118H, D1MA00538C}. These size-dependent properties play a central role in applications ranging from photonics and sensing, protection of water infrastructures \cite{apostolakis2024photoacoustic, Cousin2022} to biomedicine and energy conversion \cite{D4MA90092H,Fabrizio12, henini2011handbook,Dong:20, 11142662, Uppu2021}.
Note that NPs are not likely to replace silicon photonics entirely but are an emerging area of research that can significantly enhance and extend the capabilities of existing silicon technology, enabling new functionalities not possible with bulk silicon \cite{Zafar2025-dl,Zafarpol4,Zafarreview,Zafarpol6,zafar2023band,zafar2023compact,ZafarIEEEAccess,Zafar:22}. 
Furthermore, NPs can also potentially replace more complex and expensive superlattice and multiple quantum structures for a wealth of optolectronic devices.\cite{Pereira2020-ii,Vaks2022-uk,ma11010002,Michler1998-cp,Grempel}.

In particular, the optical response of metallic NPs, such as silver and gold, depends sensitively on the radius of the particle, which governs the spectral position, linewidth, and strength of localized surface plasmon resonances \cite{kelly2003optical}. Consequently, controlling the size and size distribution of NP remains a key requirement for applications that include surface-enhanced Raman scattering substrates, optical probes, and photothermal processes \cite{Fabrizio12, Ganguly2024, Naccache2017}. 

Despite significant progress in chemical synthesis, achieving refined and reproducible NP size control remains a challenging task. Classical approaches based on burst nucleation and growth separation, including LaMer-type mechanisms, rely primarily on chemical parameters such as precursor concentration, temperature, and reaction time \cite{Kalikmanov2013, sugimoto2019monodispersed} Although highly effective, these methods remain largely empirical and offer limited opportunities for active size tuning once nucleation has begun. This has motivated a growing interest in alternative strategies that employ external physical fields as continuous control parameters capable of modifying the nucleation landscape in situ \cite{luo2015strong, ma2022using}.

In this context, magnetic fields are particularly attractive as a result of their direct coupling to material susceptibilities, enabling them to influence thermodynamic contributions during nucleation and growth. A growing body of experimental work has demonstrated that exposure of the nucleation medium to a magnetic field can modify NP size and morphology in diamagnetic, paramagnetic, and ferromagnetic systems \cite{kthiri2021novel, luo2015strong, ualkhanova2019influence}.

Recently, we introduced an analytical framework that quantitatively links the NP size to an externally applied magnetic field by extending the classical nucleation theory (CNT) to include magnetic free-energy contributions, combined with a description of the sphere-packing of the atomic assembly within the NP \cite{Tawalbeh2025controlling}. This approach yields a closed-form relation between particle radius and magnetic field strength and reproduces the observed shift in the most-probable particle size from tens of nanometers at zero field to only a few nanometers under moderate magnetic fields \cite{kthiri2021novel}. Within the limits of classical capillarity and spherical demagnetization, the model further shows that the magnetic field lowers the nucleation barrier, effectively acting as a catalyst for nucleation.

Although atomistic approaches such as classical density functional theory and electronic density functional theory can, in principle, capture finite-size correlations and surface effects beyond the capillarity approximation, their application to NP nucleation remains computationally demanding \cite{Giovannini2023-fs, PhysRevE.98.012604, D1RA04876G}. Even the smallest NPs considered experimentally contain thousands of atoms, placing them beyond the practical limits of fully atomistic simulations. In this regime, closed-form analytical approaches offer a complementary route that preserves physical transparency while remaining quantitatively consistent with experimental trends \cite{Giovannini2023-fs}.

In the present work, we extend this framework by incorporating additional experimental observations reported in the literature, including configurations where the magnetic field orientation differs relative to the growth and stirring directions of the nucleation medium \cite{li2025effects}. These experiments show that magnetic-field-induced size reduction persists across distinct geometrical configurations and field polarizations, indicating that the underlying mechanism is robust and not restricted to a single experimental realization. By integrating these results within the same theoretical framework, we further substantiate the generality of magnetic-field-controlled NP nucleation and size selection.

\section{Methods}
\begin{figure}
    \centering
    \includegraphics[width=0.8\linewidth]{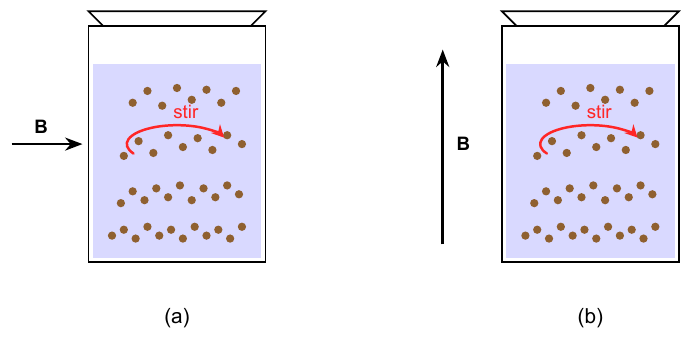}
    \caption{(a) The Group I experimental configuration where the field is parallel to the stirring direction (b) The Group II experimental configuration where the field is perpendicular to the stirring direction}
    \label{fig:1}
\end{figure}

Recent experiments on the magnetically assisted chemical reduction of AgNPs provide clear evidence that an external magnetic field acts as an effective control parameter for the size of the NP \cite{li2025effects, kthiri2021novel}. In our previous work \cite{Tawalbeh2025controlling}, we formulated a model that explains this effect for the experimental setup by Kthiri et al \cite{kthiri2021novel} which we seek to apply for the results of Li et al \cite{li2025effects}. Measurements performed under both parallel and perpendicular magnetic field configurations shown in Fig. \ref{fig:1} demonstrate a decrease in the mean particle size with increasing field strength. The data reported are summarized in table \ref{table:1}.


\begin{table}[!ht]

\centering
\label{tab:agnp_radius_nm}
\begin{tabular}{@{}ccccc@{}}
\toprule
Configuration & $B$ (mT) & $r$ (nm) & $r_{\min}$ (nm) & $r_{\max}$ (nm) \\
\midrule
 & 0.00  & 245 & 188 & 308 \\
 & 18.07 & 210 & 165 & 261 \\
Parallel & 28.65 & 180 & 146 & 218 \\
 & 38.71 & 170 & 140 & 204 \\
 & 49.27 & 170 & 141 & 205 \\
\midrule
 & 0.00   & 245 & 187 & 316 \\
 & 63.06  & 195 & 163 & 225 \\
Perpendicular & 114.21 & 175 & 144 & 204 \\
 & 157.04 & 165 & 134 & 192 \\
 & 180.78 & 155 & 129 & 183 \\
\bottomrule
\end{tabular}
\caption{AgNPs radii $r$ and corresponding size bounds as a function of magnetic field $B$ for the two experimental configurations as reported by Li et al \cite{li2025effects}}
\label{table:1}
\end{table}

Within the framework of Classical Nucleation Theory (CNT) we consider the work of formation $\Delta\mathcal{F}$ as:
\begin{equation}\label{eq:1}
    \Delta \mathcal{F} = -n(x)\bigg(\Delta \mu +\Theta(\sigma)\frac{V_{a}}{2}\frac{3}{\mu_0}\frac{|\chi_m|}{(3+\chi_m)} \mathcal{B}^2\bigg) + 4\pi r^2\gamma
\end{equation}

Here, $n$ is the number of atoms that make up a NP, which is a fit based on sphere-packing, $\Delta\mu$ is the change in the chemical potential corresponding to the phase transition, $\mathcal{B}$ is the external magnetic field measured in Tesla,  $V_a$ is the atomic volume $\frac{4}{3}\pi a^3$, $\mu_0$ is the permeability of free space and $\chi_m$ is the magnetic susceptibility. $\gamma$ is the surface free energy and $r$ is the radius of the formed NP. A more detailed discussion of the work of formation is in our previous work \cite{Tawalbeh2025controlling}. We take the average between the dipole moment and magnetic field as {$\Theta(\sigma) = |\langle\cos(\theta)\rangle|$, which is $\cos(\theta)$ weighed by a gaussian with a standard deviation $\sigma$. 

We model the number of atoms that make up an NP considering that a NP is made of a bulk part where the atoms are perfectly packed with the maximum packing density $\phi_b = \frac{\pi}{3\sqrt2}$ and a geometric shell packing is less efficient with a packing fraction $\phi_d=0.517$ as shown in Fig \ref{fig:2}. $x$ here is the reduced radius $r/a$ where $a$ is the atomic radius. The full expression is
\begin{equation}\label{eq:2}
    n(x) = \phi_b \,(x-\delta)^3 + \phi_d \,\big(x^3 - (x-\delta)^3\big)
\end{equation}

\begin{figure}[!ht]
    \centering
    \includegraphics[width=0.8\linewidth]{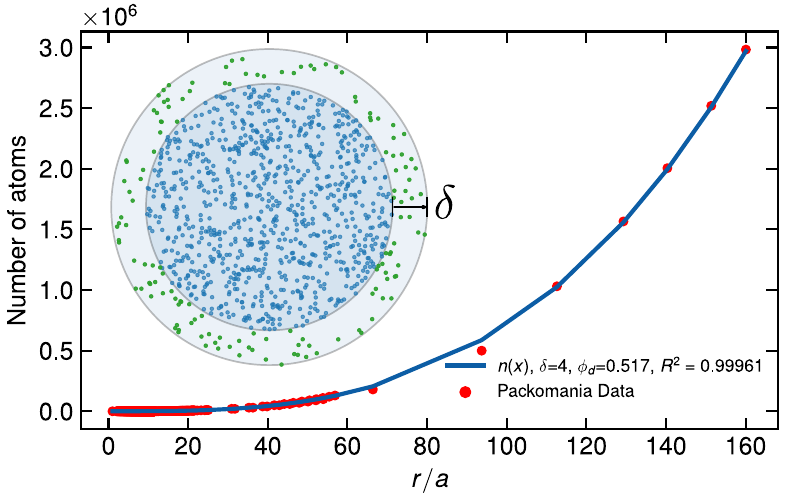}
    \caption{the sphere packing fit $n(x)$. NPs are modeled as a perfectly packed core (blue) with packing fraction $\phi_b=\frac{\pi}{3\sqrt{2}}$ and a defective surface (green) with a packing fraction $\phi_d=0.517$ obtained by fitting the data from packomania \cite{pack}. $\delta$ is the thickness of the surface.}
    \label{fig:2}
\end{figure}

In order to find a relationship between the field and the radius, we maximize $\Delta\mathcal{F}$ with respect to the radius to find \cite{Tawalbeh2025controlling} {(where $x = r/a$): 
 \begin{equation}
   \frac{x}{n'(x)}= \bigg[\frac{x_1}{n'(x_1)}+\bigg(\frac{x_1}{n'(x_1)} -\frac{x_2}{n'(x_2)}\bigg)\Theta(\sigma)\bigg(\frac{\mathcal{B}}{\mathcal{B}_2}\bigg)^2\bigg]
   \label{eq:3}
 \end{equation}
The prime indicates a derivative with respect to $x$. the pair $(x_1, \mathcal{B}_1=0)$ corresponds to the radius of the NP at $\mathcal{B}=0$, while $(x_2, \mathcal{B}_2\neq0)$ is an anchor point corresponding to the radius of the NP formed at $\mathcal{B}_2$. We solve the equation above for $x$ (or equivalently $r$) to find the full relation between $r$ and $\mathcal{B}$. A more detailed discussion of solving Eq. \eqref{eq:3} is in our previous work \cite{Tawalbeh2025controlling}. We used $\sigma = 0$ rad in this treatment, which corresponds to a complete disalignment of the magnetic moment with the magnetic field.

\section{Discussion}\label{sec12}

\begin{figure}
    \centering
    \includegraphics[width=0.5\linewidth]{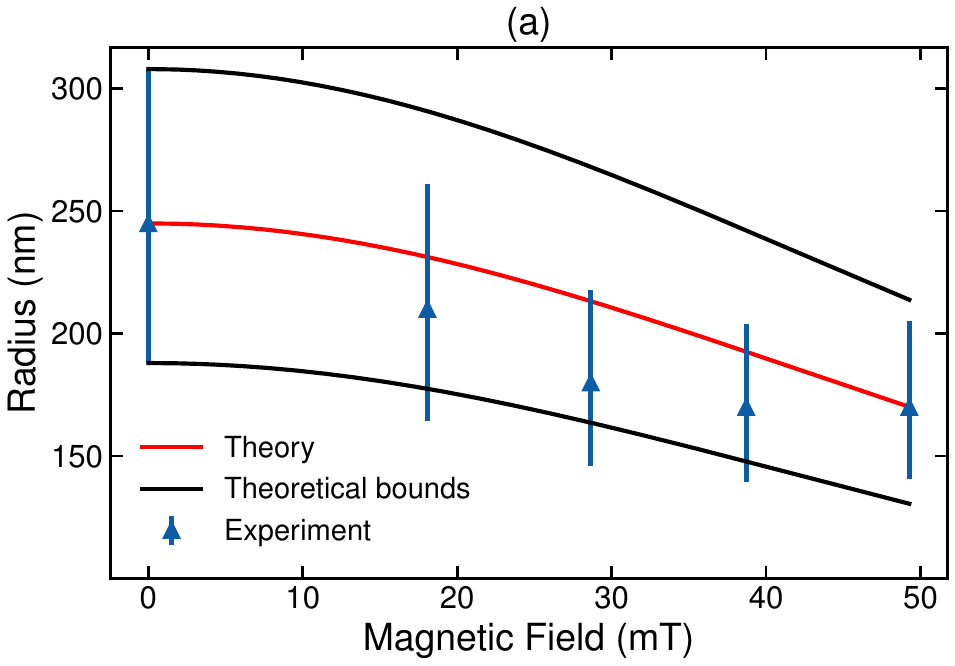}\hfill
    \includegraphics[width=0.5\linewidth]{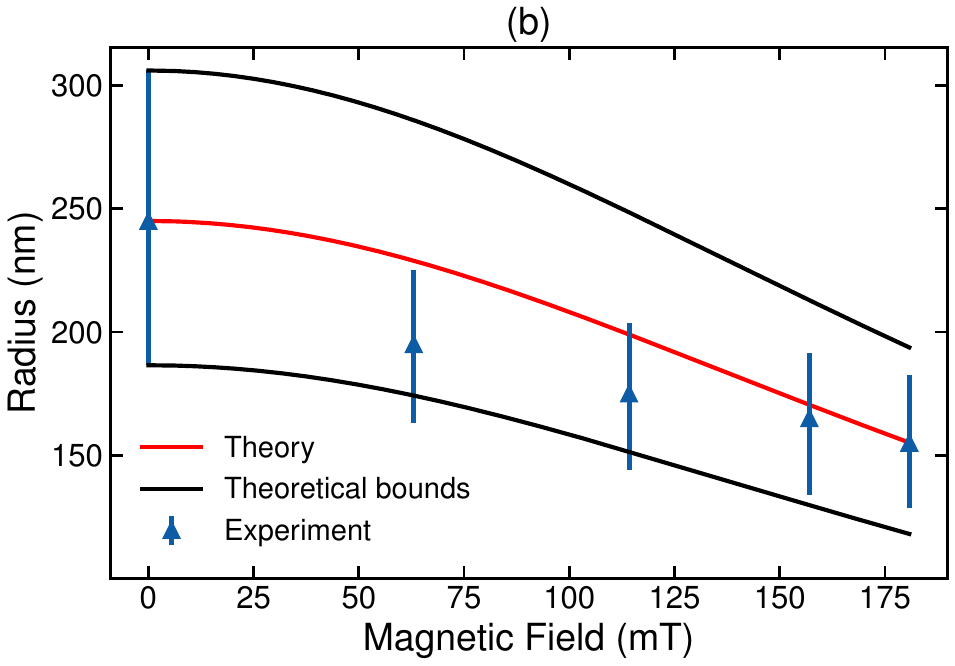}
    \caption{The field radius relation for (a) the parallel configuration and (b) the perpendicular configuration}
    \label{fig:3}
\end{figure}
Figure~\ref{fig:3} shows the dependence of the most-probable radius of AgNPs on the applied magnetic field for the parallel and perpendicular configurations. In both cases, the experimental data suggest that the AgNP radius decreases with increasing magnetic field, which is a conclusion of many previous studies \cite{kthiri2021novel, luo2015strong,ualkhanova2019influence,ma2022using}.

In the parallel configuration, the radius decreases from $r \approx 245$ nm at zero field to $r \approx 170$ nm at $\mathcal{B} \simeq 50$~mT. A comparable reduction is observed in the perpendicular configuration, though over an extended field range up to $\mathcal{B} \simeq 180$~mT. The solid curves represent the theoretical prediction obtained from our previous CNT framework introduced in~\cite{Tawalbeh2025controlling}, anchored to the zero-field radius and a single finite-field reference pointRef. The black envelopes correspond to the theoretical bounds obtained by propagating the experimental uncertainty in the zero-field radius.

In both configurations, the experimental data remain within the predicted bounds across the full field range, this confirms that the observed size reduction is a thermodynamic effect originating from the magnetic contribution to the work of formation.

The similar functional behavior observed in the parallel and perpendicular configurations indicates that the size control mechanism is robust with respect to field orientation. We note that the mechanical stirring plays a central role in particle transport and aggregation as discussed by Li et al \cite{li2025effects}. In our treatment, stirring does not directly set the critical size of the NP, but acts as a secondary transport mechanism. We capture the stirring effect The difference in the orientation of the magnetic field relative to the growth direction suggests that the magnetic field affects the surface free energy $\gamma$, which appears to be lower in the parallel orientation and leads to nucleation at a smaller critical radius. The effect of orientation on surface tension was observed experimentally by Hayakawa et al \cite{hayakawa2019effect}, where the surface tension of some liquids was lower in the orientation similar to Group I compared to the orientation of Group II. 

Although the model is formulated within CNT, its applicability is restricted to NP sizes for which capillarity-based free-energy descriptions remain meaningful; possible deviations arising from non-extensive or atomistic effects at the smallest length scales are explicitly acknowledged and discussed \cite{Tawalbeh2025controlling, MANIOTIS2025116285, Guisbiers01012019}. The NP size remains well above the size regime in which non-extensive effects dominate, further strengthening the choice of a CNT treatment. Fig. \ref{fig:4} highlights how the perpendicular orientation compares to the parallel orientation in higher fields, clearly showing the advantage that parallel orientation has over the perpendicular orientation.

\begin{figure}[!ht]
    \centering
    \includegraphics[width=.8\linewidth]{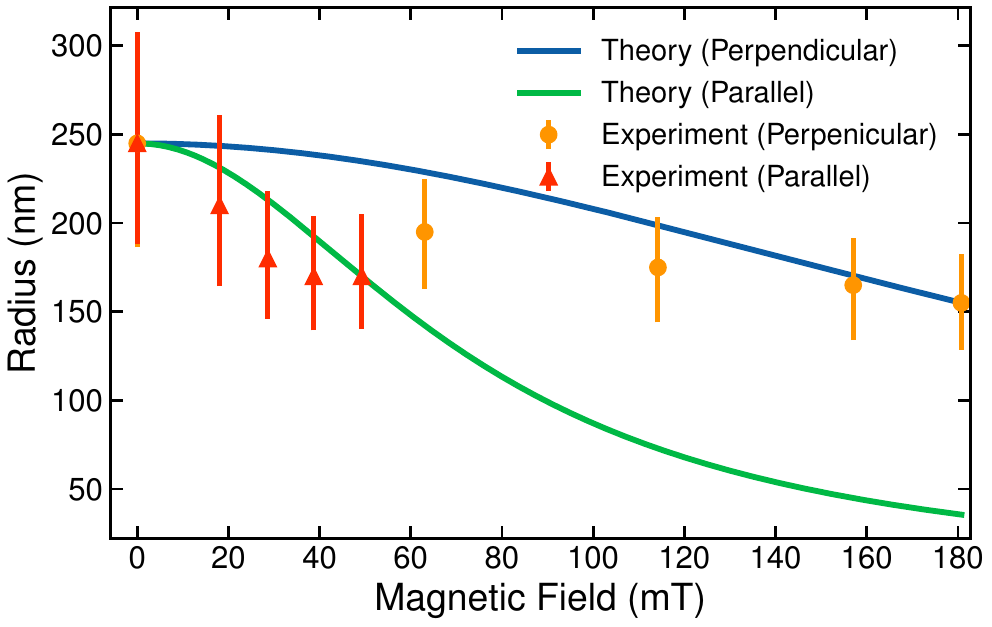}
    \caption{A comparison between the parallel and perpendicular configurations over an extended field. We can see that the radius falls off more rapidly in the parallel configuration.}
    \label{fig:4}
\end{figure}
\section{Conclusion}

We have shown that the experimentally observed reduction in AgNPs size under an external magnetic field can be consistently explained using a CNT sphere-packing approach \cite{Tawalbeh2025controlling}. When applied to the experimental data of Li et al \cite{li2025effects}, the model reproduces the experimentally observed reduction in the most–probable radius from approximately $245\,\mathrm{nm}$ at zero field to $\sim170\,\mathrm{nm}$ at $\mathcal{B}=49.27\,\mathrm{mT}$ in the parallel configuration, and to $\sim155\,\mathrm{nm}$ at $\mathcal{B}=180.78\,\mathrm{mT}$ in the perpendicular configuration. Our theory's predictions remain consistent with the experimental measurements, which strengthens our previous argument that the size reduction has thermodynamic origins. 

\backmatter

\section*{Declarations}

\begin{itemize}
\item Funding
This publication is based on work supported by Khalifa University under Award No. CIRA-2021-108.
\item Conflict of interest: The authors declare no conflict of interest.
\item Ethics approval and consent to participate: Not Applicable
\item Consent for publication: Not Applicable
\item Data availability: The datasets used and analyzed during the current
study are available from the corresponding author on reasonable
request. 
\item Materials availability: Not Applicable
\item Code availability: Not Applicable 
\item Author contribution: Mauro Pereira coordinated the driving theory project and secured funding.
Yazeed Tawalbeh extracted data from experiments in the literature and produced all the figures. Both authors contributed to
writing and reviewing the manuscript.
 \end{itemize}
 




\bibliography{sn-bibliography}

\end{document}